\documentstyle[12pt]{article}
\begin{document}

\tolerance=5000

\def\pp{{\, \mid \hskip -1.5mm =}}
\def\cL{{\cal L}}
\def\be{\begin{equation}}
\def\ee{\end{equation}}
\def\bea{\begin{eqnarray}}
\def\eea{\end{eqnarray}}
\def\tr{{\rm tr}\, }
\def\nn{\nonumber \\}
\def\e{{\rm e}}
\def\D{{D \hskip -3mm /\,}}
\def\la{\label}
\def\PLB#1 {\Jl{Phys. Lett.}{#1B}}

\  \hfill
\begin{minipage}{3.5cm}
\end{minipage}


\begin{center}
{\Large\bf Phantom and quantum matter in an Anti-de Sitter
Universe}

\vspace{6mm}

{\sc Emilio Elizalde}$^{\dagger}$\footnote{Presently on leave at
Department of Mathematics, Massachusetts Institute of Technology,
77 Massachusetts Ave, Cambridge, MA 02139. E-mail:
elizalde@math.mit.edu \ elizalde@ieec.fcr.es}
 and {\sc John Quiroga Hurtado}$^{\ddagger}$\footnote{Also at
Laboratory for Fundamental Studies, Tomsk State Pedagogical University,
634041 Tomsk, Russia. E-mail: jquiroga@utp.edu.co} \vspace{3mm}

{\sl $^\dagger$Consejo Superior de Investigaciones Cient\'{\i}ficas
(ICE/CSIC) \\  Institut d'Estudis Espacials de Catalunya (IEEC) \\
Edifici Nexus, Gran Capit\`{a} 2-4, 08034 Barcelona, Spain}
\vspace{3mm}

{\sl $^{\ddagger}$Department of Physics \\
Universidad Tecnol\'ogica de Pereira \\ Pereira, Colombia}

\vfill

{\bf Abstract}

\end{center}

We consider an Anti-de Sitter Universe filled by quantum CFT with
classical phantom matter and perfect fluid. The model represents
the combination of a trace-anomaly annihilated and a phantom
driven Anti-de Sitter Universes. The influence exerted by the
quantum effects and phantom matter on the AdS space is discussed.
Different energy conditions in this type of Universe are
investigated and compared with those for the corresponding model
in a de Sitter Universe. \vfill


\newpage

 There is growing interest towards various studies of the
reported acceleration of the scale factor of our observable
Universe \cite{SuNv}. This is motivated by recent astrophysical
data analysis, which hint to such behavior. In order to explain
it, the simplest possibility is to introduce a dark energy
component, whose origin remains however uncertain. It seems to be
possible to model the accelerating scale factor through the
introduction of phantom matter with a negative energy density
\cite{negative}. Such phantom matter may then serve as a different
possible origin for dark energy.  However, there are a number of
problems, as the violation of the energy conditions and that of a
related negative energy density, the lack of a plausible
explanation of the origin of such phantom matter, etc. Because of
those, such possibility becomes rather non-realistic.

The peculiar properties of a phantom scalar (with negative kinetic
energy) in a space with non-zero cosmological constant have been
recently discussed in an interesting paper by Gibbons
\cite{Gibbons}. It has been indicated there, that phantom
properties bear some similarity to quantum effects \cite{phtmtr}.
(Note that there are other models where the dark energy has also a
quantum origin \cite{elizalde}.) The interesting property of the
investigation in \cite{Gibbons} is that it is easily generalizable
to other constant curvature spaces, as the Anti-de Sitter (AdS)
space. There is presently some interest in such spaces, coming
specially from the AdS/CFT correspondence. According to that, the
AdS space may have in fact a cosmological influence \cite{cvetic},
increasing the number of particles created on a given subspace
\cite{jcap}. (It may also be used to study a cosmological AdS/CFT
correspondence \cite{b494}.) Hence, the study of a phantom field
in AdS space may give us a hint for the origin of such field via
the dual description. In the supergravity description, one may
think of the phantom as of a special RG flow for scalars in gauged
AdS supergravity. (Actually, such RG flow may correspond to an
imaginary scalar.)

In the present letter, we consider an AdS model filled with
classical matter, perfect fluid, and a phantom scalar, taking also
into account quantum contributions. The model can be viewed as
some generalization of phantom cosmology. In our theory, quantum
effects are described via the conformal anomaly, what is
reminiscent of the well known anomaly-driven inflation
\cite{starobinsky}. Such quantum effects are typical for the
vacuum energy (for a review, see \cite{book}). We specially
discuss the analogies between our model formulated in AdS space
and the corresponding one formulated in a de Sitter (dS) Universe
\cite{phtmtr}. Using the AdS/CFT correspondence, one may then
expect that the phantom field emerges out of some QFT instability
in the dual description. It may originate as a result of some
phase transition. We will also study how the energy conditions are
fulfilled in a phantom AdS Universe of this kind.

We start from  Einstein's equations with the scalar
(phantom) field $C$ \cite{Gibbons} \be \label{phtm1} R_{\mu\nu} -
{1 \over 2}Rg_{\mu\nu} =8\pi G \left\{ \left(\rho + p\right)U_\mu
U_\nu + p g_{\mu\nu}
 - \partial_\mu C \partial_\nu C + {1 \over 2}g_{\mu\nu} g^{\alpha\beta}
\partial_\alpha C \partial_\beta C\right\}\ .
\ee
Our model is given by 4-dimensional Anti-de Sitter spacetime
(AdS$_4$) with the metric  chosen as \cite{brevik}
\be
 ds^2=e^{-2\lambda
\tilde{x_3}}(dt^2-(dx^1)^2-(dx^2)^2)-(d\tilde{x}^3)^2\,. \la{metr}
\ee

The simplest way to account for quantum effects (at least, for
conformal matter) is to include the contributions coming from the
conformal anomaly:
\be \label{OVII} T=b\left(F+{2 \over 3}\Box
R\right) + b' G + b''\Box R\ , \ee
where $F$ is the square of 4d Weyl tensor and $G$ the Gauss-Bonnet
invariant, which are given as \bea \label{GF} F&=&{1 \over 3}R^2
-2 R_{ij}R^{ij}+ R_{ijkl}R^{ijkl}\, , \nn G&=&R^2 -4 R_{ij}R^{ij}+
R_{ijkl}R^{ijkl} \, . \eea
In general, with $N$ scalar, $N_{1/2}$
spinor, $N_1$ vector fields, $N_2$ ($=0$ or $1$) gravitons and
$N_{\rm HD}$ higher derivative conformal scalars, $b$, $b'$ and
$b''$ turn out to be \bea \label{bs} && b={N +6N_{1/2}+12N_1 + 611
N_2 - 8N_{\rm HD} \over 120(4\pi)^2}\ ,\nn && b'=-{N+11N_{1/2}+62N_1
+ 1411 N_2 -28 N_{\rm HD} \over 360(4\pi)^2}\ , \nn &&  b''=0\ .
\eea

The contributions of the conformal anomaly to $\rho$ and $p$ can be
found in \cite{NOev,NOOfrw}, namely
\bea \label{hhrA3} \rho_A&=&-\left.{1
\over a^4}\right[b'\left( 6 a^4 H^4 + 12 a^2 H^2\right)
\\
&& + \left({2 \over 3}b + b''\right)\left\{ a^4 \left( -6 H
H_{,tt}- 18 H^2 H_{,t} + 3 H_{,t}^2 \right) + 6 a^2 H^2\right\}
\nn &&  -2b +6 b' -3b'' \Bigr] ,\nn \label{hhrAA1} p_A&=&b'\left\{
6 H^4 + 8H^2 H_{,t} + {1 \over a^2}\left( 4H^2 + 8 H_{,t}\right)
\right\} \nn && \left.+ \left({2 \over 3}b + b''\right)\right\{
-2H_{,ttt} -12 H H_{,tt} - 18 H^2 H_{,t} - 9 H_{,t}^2 \nn &&
\left. + {1 \over a^2} \left( 2H^2 + 4H_{,t}\right) \right\} - {
-2b +6 b' -3b''\over 3a^4} \ . \eea

The classical matter solution corresponds to
\be \label{ph1} U_\mu = \delta^{x^3}_{\ \mu}\ ,\quad C=\tilde a
{x^3} + \tilde b\  \ee
where the latter is the solution of the equation of motion for the
phantom field:
\be \label{ph2} 0=\Box C = - \partial_{x^3}^2 C\ . \ee
Note that $\tilde a$ and $\tilde b$ are arbitrary, what implies that
there are infinitely many solutions (distributions of phantom matter) of
(\ref{ph2}).

{}For the quantum energy density and pressure, one gets
\be \label{phtm2} \rho_A=-p_A = - {6b' \lambda^4}\ . \ee
Now, since for the metric (\ref{metr}), we have
\be \label{phtm2b} R_{\mu\nu}= {-3
\lambda ^2}g_{\mu\nu}\ ,\quad R={-12 \lambda^2}\ , \ee the
$(x_3x_3)$ and $(ij)$-components in the Einstein equations
(\ref{phtm1}), with account to quantum effects are, respectively,
\bea \label{phtm3} -{3 \lambda^2}&=&8\pi G\left(\rho_{\rm matter}
- {6b' \lambda^4} - {\tilde a^2
\over 2}\right)\ ,\\
\label{phtm4}
 {3 \lambda^2}&=&8\pi G\left(p_{\rm matter} + {6b' \lambda^4} - {\tilde a^2
\over 2}\right)\ . \eea Here $\rho_{\rm matter}$ and $p_{\rm
matter}$ are contributions from the matter to the energy density
$\rho$ and pressure $p$, respectively.
By combining (\ref{phtm3}) and (\ref{phtm4}), we obtain the equations
\bea
\label{phtm5}
0&=& \rho_{\rm matter} + p_{\rm matter} - \tilde a^2 \ ,\\
\label{phtm6} 0&=& {12b' \lambda^4} + {6 \lambda^{2}\over 8\pi G }
- \rho_{\rm matter} + p_{\rm matter} \ . \eea

Let us now see what are the implications of these equations for
the different cases that arise. The first situation is in the
absence of any matter,  scalar or phantom. We easily see from
(\ref{phtm6}) that there is no solution and, therefore, the
creation of an AdS Universe is not possible. This should come as
no surprise and may be regarded as a consistency check. As a
second situation, let us consider the one when there is only QFT;
in such case we recover the same solution which was already
obtained in \cite{brevik}, i.e., the AdS Universe annihilates when
only quantum matter effects are present. As a third case, we
consider the situation when only phantom matter  is present. As we
see from (\ref{phtm6}), under this condition we do not have any
solution and this means that the creation of an AdS Universe is
not possible, as the consequence of the existence of phantom
matter only, e.g. just from the dark energy. Performing a further
analysis along the same lines, now considering the situation when
there coexist the quantum theory and phantom matter, we see again
that both the quantum matter effects and the presence of the
phantom are not enough in order to get the conditions for the
creation of an AdS Universe, either.

Consider next the more general situation where we admit in our
theory the presence of classical and quantum matter and also the
phantom field.
The solutions for Eq.(\ref{phtm6}), with respect to $\lambda^2$, are
given by the following expression
 \be \label{phtm6b} \lambda^{2}={1 \over 12b'}\left\{ -
{3 \over 4\pi G}\pm \sqrt{\left({3 \over 8\pi G}\right)^2 +
12b'\left(\rho_{\rm matter} - p_{\rm matter}\right)} \right\}\ ,
\ee
with the condition,
\be \label{phtm7} \left({3 \over 8\pi G}\right)^2 +
12b'\left(\rho_{\rm matter} - p_{\rm matter}\right)\geq 0\ . \ee
We should note that the choice of the $+$ sign (from the $\pm$)
in (\ref{phtm6b}) corresponds
to the solution of Ref. \cite{Gibbons} in the limit $b'\to 0$. On
the other hand, when $\rho_{\rm matter} - p_{\rm matter}=0$, the
solution with $-$ sign  in (\ref{phtm6b}) corresponds to Starobinsky's
anomaly-driven inflation  \cite{starobinsky}. The
limiting case $\rho_{\rm matter} = p_{\rm matter}$ describes
stiff matter. Without quantum effects, there is no non-trivial
solution for $\lambda^{-2}$ \cite{Gibbons} but, due to the
conformal anomaly, there is a nontrivial solution even for the
case of stiff matter.

With respect to the explicit Anti-de Sitter cosmological solution,
we may wonder now what kind of energy conditions can be fulfilled
in our model. The standard types of energy conditions in cosmology
are the following:
\begin{enumerate}
\item Null Energy Condition (NEC): \be \label{phtm11} \rho + p
\geq 0. \ee \item Weak Energy Condition (WEC): \be \label{phtm8}
\rho\geq 0 \ \mbox{and}\ \rho + p \geq 0. \ee \item Strong Energy
Condition (SEC): \be \label{phtm9} \rho + 3 p \geq 0\ \mbox{and}\
\rho + p \geq 0. \ee \item Dominant Energy Condition (DEC): \be
\label{phtm10} \rho\geq 0 \ \mbox{and}\ \rho \pm p \geq 0. \ee
\end{enumerate}
If we now rewrite Eqs. (\ref{phtm3}) and (\ref{phtm4}) as
\bea \label{phtm12} \rho_{\rm matter} &=& {6b' \lambda^4} +
{\tilde a^2 \over 2} - {3\lambda^2 \over 8\pi G
} ,\\
\label{phtm13} p_{\rm matter} &=& -{6b' \lambda^4} + {\tilde a^2
\over 2} + {3\lambda^2 \over 8\pi G } \ , \eea we see that in the
case of the dS Universe \cite{phtmtr}, the NEC is always satisfied
from (\ref{phtm5}): \be \label{phtm14} \rho_{\rm matter} + p_{\rm
matter}\geq 0\ . \ee The WEC could be satisfied, from
(\ref{phtm12}), if \bea \label{phtm15} && \hspace*{-25mm} {6b'
\lambda^4} + {\tilde a^2 \over 2} + {3\lambda^2 \over 8\pi G} \nn
&=& {\tilde a^2 \lambda^4\over 2}\left\{ \lambda^{-2} - {1 \over
\tilde a^2}\left( {3 \over 8\pi G} + \sqrt{\left({3 \over 8\pi
G}\right)^2 - 12b'\tilde a^2}\right)\right\} \nn && \times \left\{
\lambda^{-2} - {1 \over \tilde a^2}\left( {3 \over 8\pi G} -
\sqrt{\left({3 \over 8\pi G}\right)^2 - 12b'\tilde
a^2}\right)\right\} \nn &\geq& 0\ . \eea Since, usually $b'<0$,
the quantity inside the square root is
 positive.

{}From Eq.(\ref{phtm15}) we easily obtain a non-trivial constraint
for the length parameter $\lambda^{-2}$ \be \label{phtm16}
\lambda^{-2} \geq {1 \over \tilde a^2}\left( {3 \over 8\pi G} +
\sqrt{\left({3 \over 8\pi G}\right)^2 - 12b'\tilde a^2}\right)\ .
\ee Taking now into account that our Einstein equations differ
from those which were obtained in \cite{phtmtr}, but only in the
sign, we see that this energy condition coincides with the one
obtained in that reference for the case of the dS Universe.

In case of no quantum effects ($b'=0$), the constraint becomes
trivial: $\lambda^{-2}\geq 0$.
Since \bea \label{phtm17} \rho_{\rm matter} + 3p_{\rm matter} &=&
-{12b' \lambda^4} + 2\tilde a^2 + {3\lambda^2 \over 4\pi G} \nn
&=& {2\tilde a^2 \lambda^4}\left\{ \lambda^{-2} - {1 \over \tilde
a^2}\left( {-3 \over 16\pi G} + \sqrt{\left({3 \over 16\pi
G}\right)^2 + 6b'\tilde a^2}\right)\right\} \nn && \times \left\{
\lambda^{-2} - {1 \over \tilde a^2}\left( {-3 \over 16\pi G} -
\sqrt{\left({3 \over 16\pi G}\right)^2 + 6b'\tilde
a^2}\right)\right\}, \eea
if it turns out that
\be \label{phtm18} \left({3 \over
16\pi G}\right)^2 + 6b'\tilde a^2\leq 0 , \ee
then, the quantity inside the square root in (\ref{phtm17}) is
non-positive, and we find that the SEC is always satisfied; $\rho_{\rm
matter} + 3p_{\rm matter}\geq 0$. On the other hand, if
\be \label{phtm19} \left({3 \over 16\pi G}\right)^2 + 6b'\tilde
a^2> 0\ , \ee
then the SEC gives the following non-trivial constraint for
$\lambda^{-2}$
\bea \label{phtm20} \lambda^{-2} \geq {1 \over \tilde a^2}\left(
{-3 \over 16\pi G} + \sqrt{\left({3 \over 16\pi G}\right)^2 +
6b'\tilde a^2}\right)\ . \eea
This constraint becomes trivial ($\lambda^{-2}\geq 0$) again
if we do not include the conformal anomaly. From another side, we
observe that this constraint (\ref{phtm20}) will be always satisfied,
even when the contribution from the conformal anomaly is
included, i.e., the constraint for $\lambda^{-2}$ is always trivial.
Since $b'$ is negative for the usual matter fields,
Eq. (\ref{phtm18}) tells us that if the quantum effect is large, the
SEC could be satisfied more easily. On the other hand, if
quantum effects are small, but do not vanish, the SEC may not be
satisfied.

Concerning this energy condition, we observe that it differs from
the case of the dS Universe \cite{phtmtr}. Specifically, we see
that, since the constraint for $\lambda^{-2}$ is less strong for
the AdS space, the energy condition will be easier satisfied in
the AdS Universe than in the dS Universe.

Eq. (\ref{phtm7}) can be rewritten, for $b'<0$, as \be
\label{phtm21} \rho_{\rm matter} - p_{\rm matter} \leq - { 1\over
12b'}\left({3 \over 8\pi G}\right)^2 \ , \ee which does not
conflict with the DEC but it yields a non-trivial constraint for
the matter field. This constraint will not appear if we do not
include the contribution from the conformal anomaly. Then,
similarly to the case of the dS Universe \cite{phtmtr}, owing to
the quantum effects, there might happen that the DEC could not be
satisfied.

The contributions to $\rho$ and $p$ from the phantom field $C$
are, according to (\ref{phtm3}) and (\ref{phtm4}), given by \be
\label{phtm23} \rho_C=p_C = -{\tilde a^2 \over 2}\ . \ee Thus, we
conclude that, similarly to the case of the dS Universe, no kind
of energy conditions in the AdS Universe can be satisfied for
purely phantom matter, unless $\tilde a=0$. When $\tilde a=0$,
from (\ref{phtm5}) it follows that \be \label{phtm24} 0= \rho_{\rm
matter} + p_{\rm matter} \ , \ee which is a limiting case but does
not violate any energy condition, although its fulfillment
requires a negative pressure. We thus conclude that the energy
conditions, when quantum CFT is present, can be satisfied, unlike
what happens in the case of pure phantom matter.

To summarize, we have studied the influence of phantom and quantum
effects in an AdS Universe and drawn several interesting
conclusions. In particular, when matter is composed of phantom,
perfect fluid and quantum CFT components, we have seen that it is
sometimes possible to realize an AdS Universe, in the sense that
the majority of the energy conditions can be preserved. It would
be now interesting to investigate the cosmological implications of
a phantom field non-minimally coupled to gravity. This would be
similar to a study of the annihilation of a dilatonic AdS Universe
\cite{brevik,quiroga,quiroga1}. It would be of interest to study
the role of other phantom fields like spinors where even for usual
spinor matter there are some open questions \cite{ahluwalia}

\vspace{3mm}

\noindent {\bf Acknowledgments}

We are grateful to S.D. Odintsov for very helpful discussions. The
research of J.Q.H. (professor at UTP)has been supported by a
Professorship and Fellowship from the Universidad Tecnol\'ogica de
Pereira, Colombia. E.E. is indebted to the members of the
Mathematics Department, MIT, where this work was completed, and
specially to Dan Freedman, for warm hospitality. This
investigation has been supported by DGICYT (Spain), project
BFM2000-0810 and by CIRIT (Generalitat de Catalunya), grants
2002BEAI400019 and 2001SGR-00427.

\end{document}